# Isolator-free Integration of C-band InAs-InP Quantum Dash Buried Heterostructure Lasers with Silicon Waveguides

**Jens H. Schmid, Mohamed Rahim, Grzegorz Pakulski, Martin Vachon, Siegfried Janz, Pavel Cheben, Dan-Xia Xu, Philip J. Poole, Pedro Barrios, Weihong Jiang, Jean Lapointe, Daniele Melati**

*National Research Council Canada, Building M-50, 1200 Montreal Rd. Ottawa, Ontario, Canada, K1A-0R6*
*Jens.Schmid@nrc-cncrc.gc.ca*

**Abstract:** An InAs-on-InP quantum dash buried heterostructure laser and silicon chip optimized for mutual integration by direct facet-to-facet coupling have achieved -1.2 dB coupling efficiency, with coupled laser RIN of -150 dB/Hz and 152 kHz linewidth.

## 1. Introduction

The combination of laser sources with silicon photonic integrated circuits (PIC) has been an ongoing subject for research since first basic silicon waveguide components were proposed and demonstrated. The reported work ranges from light emitting silicon-based structures intended for monolithic integration directly with silicon waveguide circuits to heterogeneous integration methods by which light emitting III-V semiconductor lasers are either grown directly on silicon or bonded to the silicon PIC [1-3]. While the latter approaches in particular have shown promise in recent years, the packaging of separate III-V lasers and silicon chips in a hybrid assembly remains the only approach for combining III-V lasers with silicon PICs that retains the full performance advantages of both material technologies while remaining within the purview of established manufacturing methods. Hybrid assembly nevertheless also has challenges arising from the number of micro-optic components that must be assembled to sub-micrometer alignment tolerances, and the sensitivity of the lasers to reflections from other optical elements in the package. Significant assembly simplification can be achieved if both the III-V and silicon chips can incorporate new design features that can reduce the number of micro-optic components such as lenses and isolators, and minimize performance degradation due to back reflections. In this work we apply two new nanotechnology approaches on the III-V laser and silicon side that allow a laser and Si PIC to be directly coupled facet-to-facet, for devices operating in the C-band wavelength range. First, we have designed a buried-heterostructure (BH) distributed feed-back (DFB) laser that incorporates InAs on InP quantum dashes as the gain material. Although more commonly grown on GaAs substrates, InAs quantum dashes or dots grown on InP can provide gain in the C-band [4]. From an optical assembly perspective, quantum dot/dash lasers are of interest because they are predicted to have a lower linewidth enhancement factor than quantum well lasers [5] and are more inhomogeneously broadened. As a result they should be less susceptible to phase and intensity noise caused by back-reflections and mode competition. The use of quantum dashes may therefore eliminate the need for optical isolators between the silicon chip and the laser. The performance of InAs-InP based quantum dash BH lasers coupled to silicon PIC for C-band operation has not previously been reported. On the silicon chip, the use of subwavelength structured metamaterial waveguides [6] allow the designer to modify the waveguide refractive index and mode profile continuously along the propagation direction. The second innovation of this work is to use matched buried mesa laser mode converters and silicon subwavelength metamaterial couplers with near perfect mode overlap so that laser and silicon chip are coupled efficiently without intervening focusing and collimating optics.

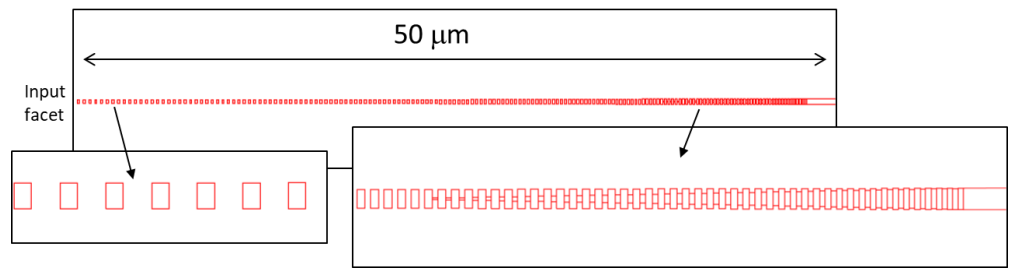

Fig. 1. Silicon metamaterial mode converter with -1 dB laser to Si facet to facet coupling loss. Insets show expanded views of facet and waveguide transition coupler sections.

## 2. Device Design and Fabrication

The BH DFB laser structure consisted of an InGaAsP alloy (1.15Q) waveguide core with three layers of InAs quantum dashes, surrounded by p- and n-doped InP. The index coupled DFB grating was formed above the waveguide core. The 1500 µm long laser waveguide was created by etching the core to form a 1350 long, 2 µm wide waveguide mesa followed by a 150 µm long mode converter that tapers to a 0.5 µm width at the facet. Selective area epitaxy was then used to add the p- and n-doped blocking layer stack that confines current to the waveguide mesa. The final p-type InP cladding and contact layers were then grown after removing the selective area masking. Both facets were anti-reflection (AR) coated. The laser has a threshold current of 14 mA, an output power of 12 mW at 100 mA, and a lasing wavelength near $\lambda=1550$ nm. The silicon chip was fabricated on a silicon-on-insulator wafer with 220 nm thick Si waveguide layer. The Si devices used in this work were 5 mm long straight waveguides with a 500 nm width covered by a $SiO_2$ cladding. At the facets the waveguides were terminated with metamaterial couplers that transition adiabatically from the 500 nm wide Si waveguide to a segmented structure at the facet. The coupler example shown in Fig. 1 has a final width of 220 nm, a 400 nm period and nominal Si/gap duty cycle of 150/400 (37.5%). Both the laser and Si mode converters were designed to produce a symmetric 3 µm wide mode profile. The measured propagation loss of the Si waveguides was 2.7 dB/cm. No AR coatings were used on the silicon facets.

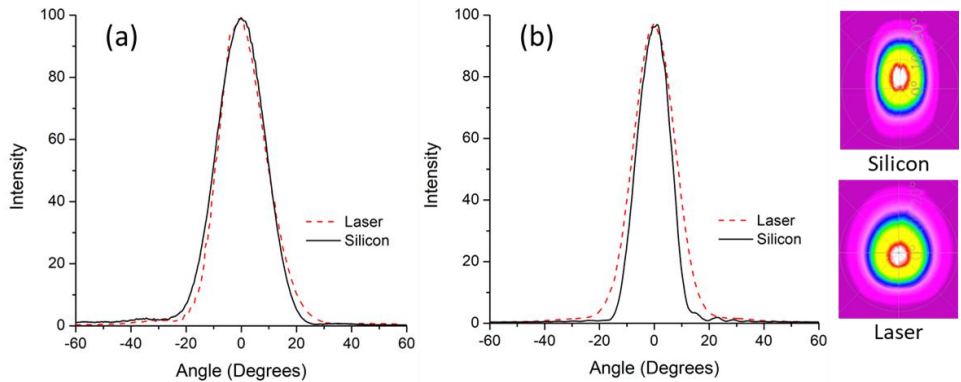

Fig. 2. Measured far field profiles for the buried heterostructure laser and Si metamaterial mode coupler for (a) the wafer normal and (b) in-plane directions. The insets show the full two-dimensional angular distributions of the far field intensity.

## 3. Experiment

The laser and Si chip were mounted on separate stages that allowed the relative separation and orientation of both to be independently adjusted, in order to emulate the possible configurations in an optical sub-assembly. The set-up allowed measurements to be carried out on individual and coupled device transmission, far field mode profiles, relative intensity noise (RIN), side-mode amplitude, and linewidth. For all measurements the laser current was set at I=100 mA. A separate fiber coupled tunable laser was used to perform stand-alone optical characterization on the Si chip waveguide. The return loss back to the laser from the silicon chip is a critical factor affecting laser performance when coupled to the Si chip. This was measured using an optical fiber coupled to the Si waveguide, and a circulator to capture the light directed backwards from the chip. The measured return loss ranged from -20 to -15 dB as the wavelength was varied. This return loss includes reflections from the Si input facet, Si output facet, light scattered backwards from sidewall roughness and also from the waveguide-coupler transitions that are not perfectly adiabatic. The rapid wavelength dependence of the return loss arises from the interference of these many contributions.

Fig. 2 shows the measured far field mode profiles for the BH quantum dash laser and the Si metamaterial couplers at $\lambda=1550$ nm. Based on the measured far field profiles, the laser and Si coupler facet mode field diameters (MFD) (at the $1/e^2$ intensity) were calculated to be $2.93 \times 2.95$ µm and $2.90 \times 3.41$ µm respectively, for the in wafer plane and out of plane directions. The coupling loss from laser to silicon chip was measured by aligning the laser to the silicon facet with a facet to facet separation of less than 1 µm. After correcting the measured transmission from the Si output waveguide for propagation loss and output facet losses, the laser to Si waveguide coupling was found to range from -1.2 dB to -2.1 dB for silicon couplers similar to that shown in Fig.1. These values include the reflection loss at the uncoated silicon waveguide facet.

The performance of the laser when coupled to the Si chip and when free standing was compared. Figure 3 shows that the side mode suppression ratio (SMSR) of the laser alone is -52 dB, increasing slightly to -48 dB when coupled to the silicon chip. The dominant side mode on the short wavelength side of the DFB grating stop band is known to be sensitive to any feedback including the residual laser facet reflection, and as a result varies from device to device even in the absence of external feedback. This is a known factor limiting manufacturing yield for uniform grating DFB lasers. Some DFB lasers sampled in this study did exhibit a poor SMSR approaching -20 dB when coupled to the Si chip. Remarkably, despite the higher SMSR in some lasers, the relative intensity noise (RIN) and laser linewidth underwent little change between coupled and uncoupled configurations for all tested quantum dash BH lasers, indicating that the quantum dash design is more immune to mode competition induced noise. The laser RIN when coupled through the Si chip was approximately -150 dB/Hz (see Fig. 3(c)), and is essentially unchanged from the RIN for the stand-alone laser. Similarly the laser linewidth, as measured with an OE Waves 4000 phase noise measurement system, was 152 kHz when transmitted through the Si chip, comparable to the 146 kHz linewidth obtained for the laser coupled directly to the instrument. Quantum well lasers coupled to Si under the same conditions in our lab displayed lasing wavelength and side mode fluctuation, along with a deterioration of RIN.

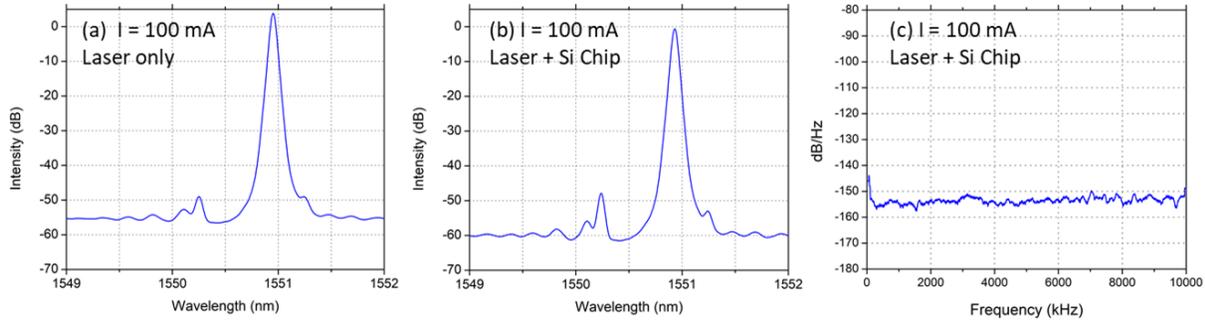

Fig. 3. Laser spectra measured (a) directly from the laser facet and (b) after transmission through the Si waveguide. (c) The relative intensity noise measured after coupling through the Si chip.

## 4. Summary

We have designed and tested InAs on InP quantum dash buried heterostructure DFB lasers and silicon photonic chips with matched mode converters to facilitate direct coupling of the laser and chip with no intervening optics. Laser to Si waveguide coupling losses as low as -1.2 dB were measured. As expected for a uniform grating DFB laser, the SMSR is sensitive to return loss, but a coupled SMSR of -48 dB was still achieved. When coupled to the silicon chip the observed RIN and laser linewidth near -150 dB/Hz and 150 kHz respectively were unchanged relative to measurements carried out on the laser alone, despite a return loss from the Si chip of up to -15 dB. This is consistent with previous theoretical and experimental work suggesting that quantum dot/dash lasers have a high tolerance to feedback and mode competition, and are hence attractive for hybrid integration. This work demonstrates that hybrid assembly can be simplified and performance improved by applying recent innovations in the nanoscale manipulation of semiconductor material properties using both photonic metamaterials and quantum dash gain media.